\documentclass[twocolumn,prb,superscriptaddress]{revtex4}
\usepackage{amsfonts}
\usepackage[T1]{fontenc}
\usepackage{amsmath,amsbsy,amssymb,graphicx}
\usepackage{times}
\usepackage{amsmath}
\usepackage{amssymb}
\usepackage{graphicx}

\let\mathbf=\boldsymbol
\begin{document}

\title{Fermion Fractionalization to Majorana Fermions in Dimerized Kitaev Superconductor}



\author{Ryohei Wakatsuki}
\affiliation{Department of Applied Physics, University of Tokyo, Hongo 7-3-1, 113-8656, Japan}
\author{Motohiko Ezawa}
\affiliation{Department of Applied Physics, University of Tokyo, Hongo 7-3-1, 113-8656, Japan}
\author{Yukio Tanaka}
\affiliation{Department of Applied Physics, Nagoya University, Nagoya 464-8603, Japan}
\author{Naoto Nagaosa}
\affiliation{RIKEN Center for Emergent Matter Science (CEMS), Wako 351-0198, Japan}
\affiliation{Department of Applied Physics, University of Tokyo, Hongo 7-3-1, 113-8656, Japan}

\begin{abstract}
We study theoretically a one-dimensional dimerized Kitaev superconductor model which belongs to BDI class with time-reversal, particle-hole, and chiral symmetries.
There are two sources of the particle-hole symmetry, i.e., the sublattice symmetry and superconductivity.
Accordingly, we define two types of topological numbers with respect to the chiral indices of normal and Majorana fermions, which offers an ideal laboratory to examine the interference between the two different physics within the same symmetry class.
Phase diagram, zero-energy bound states, and conductance at normal metal/superconductor junction of this model are unveiled from this viewpoint.
Especially, the electron fractionalization to the Majorana fermions showing the splitting of the local density of states is realized at the soliton of the dimerization in this model.
\end{abstract}

\maketitle

\address{{\normalsize Department of Applied Physics, University of Tokyo, Hongo 7-3-1, 113-8656, Japan }}

\section{Introduction}

Classification of the gapped electronic states from the viewpoint of quantum topology has shed a new light on our understanding of the physical properties of solids.
Topological insulators and superconductors are the two major ingredients of this classification \cite{Hasan,Qi,TSN}.
The topological periodic table has been proposed based on the time-reversal, particle-hole, and chiral symmetries, which are the three fundamental and robust symmetries of the Hamiltonian even without the translational symmetries or point-group symmetries \cite{Ryu,AZ,Schnyder,KitaevP}.
10 classes are identified in this table, and the homotopy group is allocated to each class depending on the spatial dimensionality of the system.
This mathematical classification alone, however, does not provide the physical properties of the concrete systems, nor provide the way how to construct the topological indices linked to the zero-energy bound states at the boundary of the sample.
Therefore, the studies of explicit models are needed to explore the rich physics hidden in the periodic table.
One interesting question is how the two different physical phenomena, characterized by each topological index, are related within the same symmetry class.
To examine this question in the simplest model, we analyze in this paper the dimerized Kitaev model, which belongs to the BDI class and is a hybrid system comprised of the spinless Su-Schrieffer-Heeger model of polyacetylene \cite{SSH,TLM,Niemi,JR} and the Kitaev model of the one-dimensional (1D) $p$-wave topological superconductor \cite{Kitaev,Alicea,Flen,Beenakker}.

The Su-Schrieffer-Heeger (SSH) model, i.e., dimerized one-dimensional chain, is a model proposed for polyacetylene.
At the edges of the sample or at the kink of the dimerization pattern, i.e., soliton, the zero-energy in-gap bound states appear due to the topological reason.
On the other hand, the Kitaev model is the one-dimensional spinless $p$-wave topological superconductor, where superconducting pairing occurs between the nearest-neighbor sites.
The finite chain of the Kitaev model supports Majorana fermions at edges.
The Kitaev model is realized by using a 1D nanowire with strong Rashba spin-orbit interaction \cite{FuKane,Sau,Alicea2,Sato2,Oreg,Lutchyn,Stanescu}. Several experiments about the nanowire systems have so far been reported \cite{Mourik,Das,Deng,Rokhinson,EJHLee1,EJHLee2}.
Both models have the particle-hole symmetry.
However, their origins are different.
In the case of the SSH model, the sublattice symmetry between the $A$ and $B$ sublattices gives the particle-hole symmetry, while the superconductivity is the source in the Kitaev model.
Correspondingly, we can define the two kinds of topological indices, $N_1$ and $N_2$ in Eqs. (\ref{N1}) and (\ref{N2}), respectively.
$N_1$ is induced by the sublattice symmetry and equals the number of zero-energy states, while $N_2$ is induced by the particle-hole symmetry due to the superconductor and equals the number of Majorana zero-energy states.
By these two indices, the phase diagram is determined, and their relation to the zero-energy states at the edges and the associated transport properties are revealed.
We also investigate the zero-energy states in the presence of a dimerization soliton in our hybrid model.
As is expected, a zero-energy fermionic state appears in the SSH-like region, which is eventually suppressed by the $p$-wave pairing.
Remarkably, we find a peak of the local density of states (LDOS) at zero energy splits into two peaks which shift toward the edges by the effect of the $p$-wave pairing.
It is regarded as a precursor of the topological phase transition, where one fermion at the soliton splits into two Majorana fermions.
This offers a unique opportunity to see the process of electron fractionalization in the real space.

The rest of the paper is organized as follows.
In Sec. \ref{sec:Hamiltonian}, we introduce the model and derive the energy bands of the bulk system.
In Sec. \ref{sec:Symmetry}, we discuss the symmetry of the model.
We show that there are two particle-hole symmetries in the system.
In Sec. \ref{sec:mu_zero}, we focus on the sublattice symmetric case ($\mu=0$).
We calculate the topological number induced from the sublattice symmetry.
In Sec. \ref{sec:mu_nonzero}, we consider the case of sublattice asymmetric case ($\mu\not=0$).
We define and calculate another topological number induced from the superconductivity.
In Sec. \ref{sec:ene}, we illustrate the energy spectrum of a finite system.
We show the number of zero-energy states reflects the two topological indices.
In Sec. \ref{sec:cnd}, we calculate the differential conductance and show the correspondence between the topological index from the superconducting pairing.
In Sec. \ref{sec:ldos}, we study the LDOS at zero energy.
We explain how edge states and soliton states appear depending on the phase.
We also derive the continuum model and show the analytic form of the zero-energy wave function or local density of states in good agreement with the numerical results.
We report our remarkable finding of the splitting of a peak in the LDOS in the presence of a soliton due to the superconducting pairing.
In Sec. \ref{sec:odd}, we discuss the relationship between the odd-frequency pairing and the soliton states.
In Sec. \ref{sec:conclusion}, we summarize the results of this paper and briefly discuss the relevance to the real systems.

\begin{figure}[t]
\centerline{\includegraphics[width=0.45\textwidth]{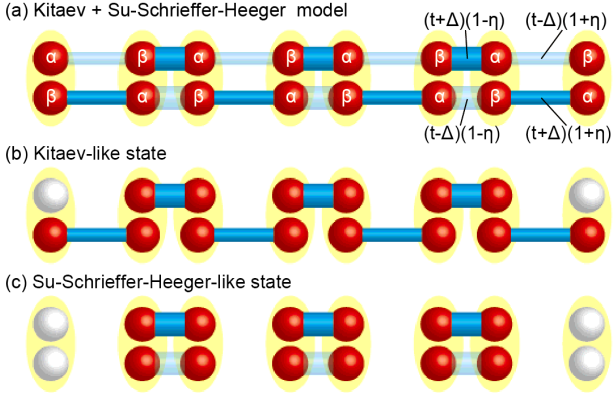}}
\caption{(Color online) (a) Illustration of the model. The red spheres, the yellow ovals, and the blue sticks represent Majorana fermions, ordinary fermions, and bonds between Majorana fermions, respectively. $\alpha$ and $\beta$ on the red spheres denote the Majorana operators which are defined in Eq. (\ref{MF}). The dark and light bonds represent dimerization of Majorana fermions due to the $p$-wave pairing. The thick and thin bonds represent the dimerization of the ordinary fermions. The coupling parameters are shown. Here, we assumed $\mu=0$ and $\eta < 0$. (b) In the Kitaev-like phase, the dark bonds are dominant. There are unpaired Majorana fermions. (c) In the SSH-like phase, the thick bonds are dominant. There are two fermions.}
\label{KSSH_illust}
\end{figure}

\section{Hamiltonian} \label{sec:Hamiltonian}

We investigate the tight-binding model for a hybrid system comprised of the SSH model \cite{SSH} and the Kitaev model \cite{Kitaev}:
\begin{align}
H = & - \mu \sum_{j}(c_{A,j}^{\dagger }c_{A,j}+c_{B,j}^{\dagger}c_{B,j}) \notag \\
& - t \sum_j \left[ \left( 1+\eta \right) c_{B,j}^{\dagger}c_{A,j}+\left( 1-\eta \right) c_{A,j+1}^{\dagger }c_{B,j}+\text{H.c.} \right]
\notag \\
& + \Delta \sum_j \left[ \left( 1+\eta \right) c_{B,j}^\dagger c_{A,j}^\dagger + \left( 1-\eta \right) c_{A,j+1}^\dagger c_{B,j}^\dagger+\text{H.c.} \right],
\label{KSSHHamiltonian}
\end{align}
where $A$ and $B$ denote the sublattice indices, $\mu$ is the chemical potential, $t$ is the transfer integral, and $\Delta $ is the superconducting pairing gap taken to be real.
The space-dependent variable of the SSH model is the dimerization $\eta$, which we have taken to be a constant for the ground state.
It contributes to the transfer integral and the superconducting pairing.
The Hamiltonian (\ref{KSSHHamiltonian}) is reduced to the Kitaev model for $\eta=0$ and to the SSH model for $\mu=0, \Delta=0$.
There is a condition on the dimerization, $|\eta|<1$, since the transfer integral should be positive.
We also assume $|\Delta| < t$.
We shall later investigate the system in the presence of the soliton excitation in the SSH model.
We show the illustration of the model in Fig. \ref{KSSH_illust}.

Introducing the four-component operator $C_{k}^{\dagger }=(c_{kA}^{\dagger},
c_{kB}^{\dagger },c_{-kA},c_{-kB})$, we can express the Hamiltonian $H$ in the Bogoliubov--de Gennes form.
In the momentum space, it reads as
\begin{equation}
H= \frac{1}{2} \sum_k C_{k}^{\dagger } \mathcal{H} \left( k \right) C_{k}
\end{equation}
with
\begin{equation}
\mathcal{H} \left( k \right)=
\begin{pmatrix}
-\mu & z & 0 & w \\
z^* & -\mu & -w^* & 0 \\
0 & -w & \mu & -z \\
w^* & 0 & -z^* & \mu
\end{pmatrix}
,
\end{equation}
where
\begin{align}
z \left( k \right) &=- t \left[ \left( 1+\eta \right) + \left( 1-\eta \right) e^{-ika} \right], \label{EqZ}\\
w \left( k \right) &=- \Delta \left[ \left( 1+\eta \right) - \left( 1-\eta \right) e^{-ika} \right], \label{EqW}
\end{align}
and $a$ is the lattice constant.
We diagonalize the Hamiltonian and obtain the eigenvalues,
\begin{align}
E^2=\mu^2 + |z|^2 +|w|^2 \pm 2 \sqrt{ \mu^2 |z|^2 + \left( 4 t \Delta \eta \right)^2}
\end{align}
with
\begin{align}
&| z \left( k \right) |^2 =2 t^{2} \left[ \left( 1+\eta^2 \right) + \left( 1-\eta^2 \right) \cos k a \right], \\
&| w \left( k \right) |^2 =2 \Delta^2 \left[ \left( 1+\eta^2 \right) - \left( 1-\eta^2 \right) \cos k a \right].
\end{align}
We find
\begin{equation}
E(0)=\pm 2t \pm \sqrt{ \mu^2 + 4 \Delta^2 \eta^2},
\end{equation}
where the gap closes at
\begin{equation}
\mu^2 = 4 \left( t^2 - \Delta^2 \eta^2 \right),  \label{gapclose_zero}
\end{equation}
while we find
\begin{equation}
E(\pi / a)=\pm 2 t \eta \pm \sqrt{\mu^2 + 4 \Delta^2},
\end{equation}
where the gap closes at
\begin{equation}
\mu^2 = 4 \left( t^2 \eta^2 - \Delta^2 \right). \label{gapclose_pi}
\end{equation}
We will show that gap-closing conditions (\ref{gapclose_zero}) and (\ref{gapclose_pi}) correspond to the phase boundaries.

For $\Delta=0$, the energy spectrum is reduced to that of the SSH model,
\begin{equation}
E(k)=\pm \mu \pm t \sqrt{2 \left[ \left( 1+\eta^2 \right) + \left( 1-\eta^2 \right) \cos k a \right]}.
\end{equation}
It is well known \cite{Ryu,AZ,Schnyder,KitaevP} that the system is topological for $\eta <0$ and trivial for $\eta >0$.

On the other hand, for $\eta=0$, the energy spectrum is reduced to that of the Kitaev model,
\begin{equation}
E(k)=\pm \sqrt{(2t\cos \frac{k a}{2}-\mu )^{2}+4\Delta ^{2}\sin ^{2}\frac{k a}{2}}.
\end{equation}
It is well known \cite{Kitaev,Ryu,AZ,Schnyder,KitaevP} that the system is topological for $|\mu|<2t$ and trivial for $|\mu|>2t$.

\section{Symmetry} \label{sec:Symmetry}

We discuss the topological class of the model.
In this spinless system, the time-reversal operator is defined by $T= K$, which takes the complex conjugate.
The model has the time-reversal symmetry $T \mathcal{H} \left( k \right) T^{-1} = \mathcal{H} \left( -k \right)$ because there is no complex coefficient of $\mu, t, \Delta$ and $\eta$ in the Hamiltonian.
It is noted that we have chosen the gauge of real $\Delta$ in Eq. (\ref{KSSHHamiltonian}).
Moreover, in the case of $\mu=0$, the system has the sublattice symmetry.
The sublattice symmetry operator is defined by
\begin{equation}
C_1 = \sigma_z =
\begin{pmatrix}
1 & 0 & 0 & 0 \\
0 & -1 & 0 & 0 \\
0 & 0 & 1 & 0 \\
0 & 0 & 0 & -1
\end{pmatrix}
,
\end{equation}
where $\sigma_i$ is the Pauli matrix acting on the sublattice degree of freedom.
It is checked that $C_1 \mathcal{H} \left( k \right) C_1^{-1} = -\mathcal{H} \left( k \right)$.
The topological class is BDI since $T^2=1$ and $C_1^2=1$.

On the other hand, if $\mu$ is finite, there is no sublattice symmetry anymore.
However, the class is still BDI due to the particle-hole symmetry of the superconductor.
The particle-hole operator is defined by $P=\tau_x K$, where $\tau_i$ is the Pauli matrix acting on the particle-hole space.
We can check that the Hamiltonian satisfies $P \mathcal{H} \left( k \right) P^{-1}= - \mathcal{H} \left( -k \right)$.
Then, the chiral operator is induced as the product of the time-reversal operator $T$ and the particle-hole operator $P$:
\begin{equation}
C_2 =T P= \tau_x =
\begin{pmatrix}
0 & 0 & 1 & 0 \\
0 & 0 & 0 & 1 \\
1 & 0 & 0 & 0 \\
0 & 1 & 0 & 0
\end{pmatrix}
.
\end{equation}
It is checked that $C_2 \mathcal{H} \left( k \right) C_2^{-1} = - \mathcal{H} \left( k \right)$ and $C_2^2=1$.
Therefore, the topological class is BDI.

The 1-D system in the BDI class is characterized by the $\mathbb{Z}$-index.
We will show soon that these two chiral operators induce the two topological $\mathbb{Z}$-indices in the case of $\mu=0$.

\begin{figure}[t]
\centerline{\includegraphics[width=0.5\textwidth]{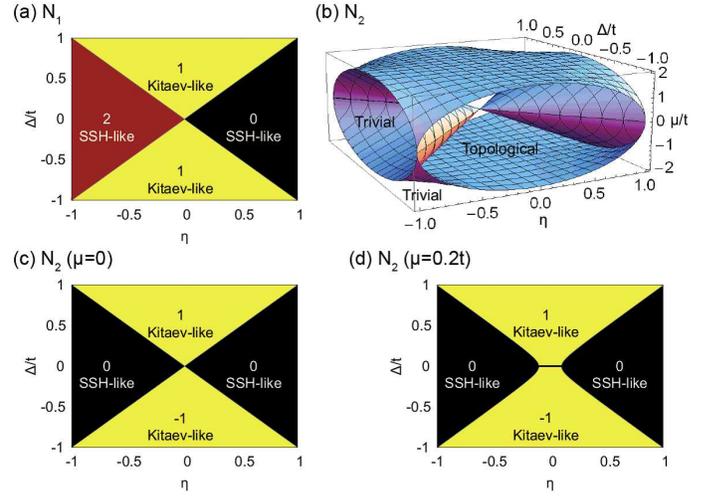}}
\caption{(Color online) (a) Topological phase diagram with respect to $N_1$ ($\mu=0$ case). The horizontal axis is $\eta$ and the vertical axis is $\Delta/t$. The numbers in the figure denote $N_1$. (b) Topological phase diagram with respect to $N_2$. The axes are $\eta, \Delta/t$, and $\mu/t$. In the trivial regions, $N_2=0$. In the topological region, $N_2=\pm 1$, depending on the sign of $\Delta/t$. The gap-less phase $\Delta/t=0$ in the Kitaev-like phase is not illustrated for the sake of clarity. (c) The cross section of (b) at $\mu=0$. (d) The cross section of (b) at $\mu=0.2t$. The black line is the gapless phase.}
\label{PD}
\end{figure}

\section{Sublattice symmetric case} \label{sec:mu_zero}

We start with the case of $\mu=0$.
First we examine the gap-closing condition.
The eigenvalues are%
\begin{equation}
E(k) = \pm 2 \sqrt{ \left( t \pm \Delta \eta \right)^2 \cos ^2 \frac{k a}{2} + \left( t \eta \pm \Delta \right)^2 \sin ^2 \frac{k a}{2} }.
\end{equation}%
It vanishes at $k=0$, $\Delta \eta = \pm t$ and $k=\pi / a$, $t \eta = \pm \Delta$.
However, since $|\Delta| |\eta|<t$, the gap closes at the points $k=\pi / a$, $t \eta = \pm \Delta$.
As we shall soon show, the topological phase boundary is given by this gap-closing condition.

The topological number associated with the sublattice symmetry operator $C_1$ is defined by
\begin{equation}
N_{1}=\text{Tr} \int_{-\pi / a}^{\pi / a} \frac{dk}{4\pi i} C_1 g^{-1} \partial_k g, \label{N1}
\end{equation}
where $g \left( k \right)=-\mathcal{H}^{-1} \left( k \right)$ is the Green's function at zero energy \cite{Gurarie,Manmana,Wang1,Wang2}.

This topological number is equivalent to the chiral index.
We introduce a unitary transformation,
\begin{equation}
U_1 =
\begin{pmatrix}
1 & 0 & 0 & 0 \\
0 & 0 & 1 & 0 \\
0 & 1 & 0 & 0 \\
0 & 0 & 0 & 1
\end{pmatrix}
,
\end{equation}
which yields
\begin{equation}
U_1 C_1 U_1^\dagger = \tau_z, \quad U_1 \mathcal{H} U_1^\dagger =
\begin{pmatrix}
0 & V_1 \\
V_1^\dagger & 0
\end{pmatrix}\label{EqHamil}
,
\end{equation}
with
\begin{equation}
V_1=
\begin{pmatrix}
z & w \\
-w & -z
\end{pmatrix},
\end{equation}
where $z$ and $w$ are defined by Eqs. (\ref{EqZ}) and (\ref{EqW}).
When the Hamiltonian is in the form of Eq. (\ref{EqHamil}), the chiral index is given by
\begin{align}
N_{1}&=-\text{Tr}\int_{-\pi / a}^{\pi / a} \frac{dk}{2\pi i} V_1^{-1} \partial_k V_1
=-\int_{-\pi / a}^{\pi / a} \frac{dk}{2\pi i} \partial_k \log \text{Det} V_1 \notag \\
&=-\sum_{n=1,2} \int_{-\pi / a}^{\pi / a} \frac{dk}{2\pi i} \partial_k \log z_n(k). \label{EqN1}
\end{align}
with%
\begin{align}
z_1 (k) &= \left( t-\Delta \right) \left( 1+\eta \right) + \left( t+\Delta \right) \left( 1-\eta \right) e^{-ika}, \\
z_2 (k) &= \left( t+\Delta \right) \left( 1+\eta \right) + \left( t-\Delta \right) \left( 1-\eta \right) e^{-ika}.
\end{align}
It is straightforward to derive that
\begin{equation}
N_{1} = \Theta \left( \Delta - t \eta \right) + \Theta \left( -\Delta - t \eta \right), \label{TopoN1}
\end{equation}
with
\begin{equation}
\Theta (x) =
\begin{cases}
0 \quad (x<0), \\
1 \quad (x>0).
\end{cases}
\end{equation}
Clearly, $N_1$ is the winding number of $z_n(k)$.
Its mathematical meaning is that $\pi _{1}\left( GL\left(4,\mathbb{C}\right) \right) =\mathbb{Z}$.

We may derive the phase diagram from Eq. (\ref{TopoN1}), as illustrated in Fig. \ref{PD}(a).
We find three phases:\newline
(i) $t|\eta| > |\Delta|, \eta > 0$, where $N_1=0$ (SSH-like trivial);\newline
(ii) $t|\eta| > |\Delta|, \eta < 0$, where $N_1=2$ (SSH-like topological); \newline
(iii) $t|\eta| < |\Delta|$, where $N_1=1$ (Kitaev-like topological).\newline
The dimerization and the $p$-wave pairing compete and result in these phases.

\section{Sublattice Asymmetric Case} \label{sec:mu_nonzero}

We proceed to investigate the $\mu \ne 0$ case.
The topological number associated with the chiral operator $C_2$ is defined by
\begin{equation}
N_{2}=\text{Tr} \int_{-\pi / a}^{\pi / a} \frac{dk}{4 \pi i} C_2 g^{-1} \partial_k g. \label{N2}
\end{equation}
This topological number is identical to the chiral index of Majorana fermion \cite{Sato,Tewari}.

We consider a unitary transformation,
\begin{equation}
U_2=\frac{1}{\sqrt{2}}
\begin{pmatrix}
1 & 0 & 1 & 0 \\
0 & 1 & 0 & 1 \\
-i & 0 & i & 0 \\
0 & -i & 0 & i
\end{pmatrix}
,
\end{equation}
which corresponds to the representation with the Majorana operators:
\begin{equation}
c_{i}=\frac{1}{2}(\alpha_i+i\beta_i),\quad c_{i}^{\dagger }=\frac{1}{2} (\alpha_i-i\beta_i). \label{MF}
\end{equation}
It follows that
\begin{equation}
U_2 C_2 U_2^\dagger = \tau_z, \quad U_2 H U_2^\dagger  =
\begin{pmatrix}
0 & V_2 \\
V_2^{\dagger } & 0
\end{pmatrix}
,
\end{equation}
with
\begin{equation}
V_2=
\begin{pmatrix}
- i \mu & i\left( z-w \right) \\
i\left( z^*+w ^*\right) & - i \mu
\end{pmatrix}
.
\end{equation}
The chiral index is given by a formula similar to Eq. (\ref{EqN1}) with the use of $V_2$ in place of $V_1$,
\begin{align}
N_{2}
=-\text{Tr}\int_{-\pi / a}^{\pi / a} \frac{dk}{2\pi i} V_2^{-1}\partial _{k}V_2 = -\int_{-\pi / a}^{\pi / a} \frac{dk}{2\pi i} \partial _{k}\log Z \left(k\right),
\end{align}
where
\begin{align}
Z\left( k\right) &=\text{Det}V_2\left( k\right)
= -\mu^2 + \left( -z+w \right) \left( -z^*-w^* \right) \notag \\
&= -\mu^2 + 2 \left( t^2-\Delta^2 \right) \left( 1+\eta^2 \right) \notag \\
& \quad + 2 \left( t^2+\Delta^2 \right) \left( 1-\eta^2 \right) \cos k a - 4 i t \Delta \left( 1-\eta^2 \right) \sin k a.
\end{align}
$N_2$ is the winding number of $Z(k)$, and determined by the cross points of the real axis at $k=0$ and $\pi / a$.
For $\Delta > 0$, we find
\begin{align}
& Z\left( 0 \right) Z\left( \pi / a \right) < 0 \Rightarrow N_2=1, \\
& Z\left( 0 \right) Z\left( \pi / a \right) > 0 \Rightarrow N_2=0,
\end{align}
with
\begin{align}
Z\left( 0\right) &=-\mu^2 + 4 \left( t^2 - \Delta^2 \eta^2 \right), \\
Z\left( \pi / a\right) &=-\mu^2 + 4\left( t^2 \eta^2 - \Delta^2 \right).
\end{align}
For $\Delta < 0$, we find $N_2=-1$ in the topological region.
However, the sign of $N_2$ is meaningless because it depends on the choice of the global phase.
The relative sign of $N_2$, on the other hand, matters when two superconductors are attached.
The phase diagram for $N_2$ is shown in Figs. \ref{PD}(b)--\ref{PD}(d).
The gap closes at the phase boundary, that is,
$Z(0)=0$ at $\mu^2 = 4 \left( t^2 - \Delta^2 \eta^2 \right)$, and
$Z(\pi / a)=0$ at $\mu^2 = 4\left( t^2 \eta^2 - \Delta^2 \right)$,
in consistent with Eqs. (\ref{gapclose_zero}) and (\ref{gapclose_pi}).

\section{Finite Chain} \label{sec:ene}

\begin{figure}[t]
\centerline{\includegraphics[width=0.5\textwidth]{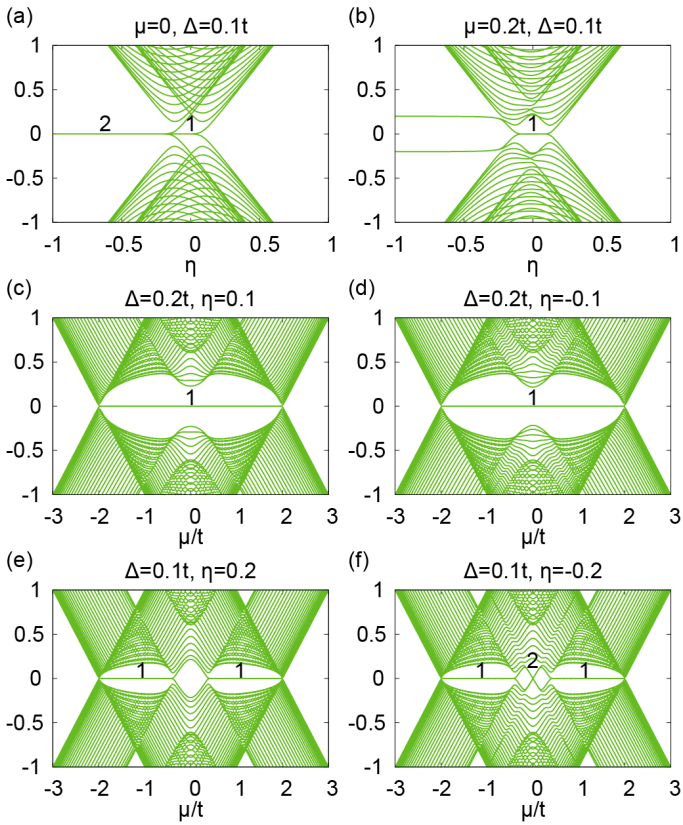}}
\caption{(Color online) Energy spectrum of the finite chain as a function of (a), (b) $\eta$, and (c)--(f) $\mu/t$. The numbers in the figures denote the degeneracy of zero-energy states divided by 2. Namely, ``1'' means one pair of Majorana fermions, and ``2'' means two fermions. These states are localized at the edges as in Fig. \ref{LDOS}. We have taken $L=64$.}
\label{ene}
\end{figure}

It is an interesting problem as to how the energy spectrum changes in the SSH model when the $p$-wave superconducting pairing is introduced.
We show the energy spectrum of the finite system as a function of $\eta$ in Figs. \ref{ene}(a) and \ref{ene}(b), where we have set $\mu =0, 0.2t$.
Without the superconducting pairing, there are only two phases, i.e., trivial for $\eta>0$ and topological for $\eta<0$, where two zero-energy fermions exist at the ends.
In the presence of the superconducting pairing, the third phase emerges for $t |\eta| < \sqrt{\left( \mu/2 \right)^2 + \Delta^2}$.
It is the Kitaev-like topological phase, where there exists one pair of Majorana fermions.
This can also be confirmed in Fig. \ref{ene}(b), where the zero-energy states in the Kitaev-like region remain while the SSH-like zero-energy states split with finite but small $\mu$.

Next we investigate how the energy spectrum changes in the Kitaev model when the dimerization is included,
which we illustrate for four sets of $\Delta$ and $\eta$ in Figs. \ref{ene}(c)-\ref{ene}(f).
Without the dimerization, the system is topological for $|\mu| < 2 t$, where there is one pair of Majorana fermions.
In the case of $\Delta > t |\eta|$ [Figs. \ref{ene}(c) and \ref{ene}(d)], the system is in the Kitaev-like phase,
where one pair of Majorana fermions appears for $|\mu| < 2 \sqrt{t^2-\Delta^2 \eta^2}$ irrespective of the sign of $\eta$.
Namely, the Kitaev-like topological phase is suppressed by the dimerization.
In the case of $\Delta < t |\eta|$ [Figs. \ref{ene}(e) and \ref{ene}(f)], the system is in the SSH-like phase for small $\mu$.
Especially, when $\mu$ is zero, the sublattice symmetry exists and there are two fermions at the edges with negative $\eta$.
For $2 \sqrt{t^2 \eta^2 - \Delta^2} < |\mu| < 2 \sqrt{t^2-\Delta^2 \eta^2}$, the system belongs to the Kitaev-like phase and supports one pair of Majorana fermions.

\begin{figure}[t]
\centerline{\includegraphics[width=0.5\textwidth]{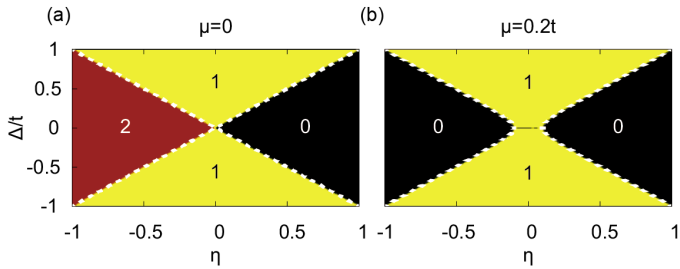}}
\caption{(Color online) The number of zero-energy states as a function of  $\eta$ and $\Delta/t$ for (a) sublattice symmetric case ($\mu=0$), and (b) sublattice asymmetric case ($\mu = 0.2t$). The white broken lines denote the phase boundaries. We have taken $L=512$ for the calculation.}
\label{zes}
\end{figure}

We also show the number of zero-energy states in Fig. \ref{zes}.
The number is equal to $N_1$ for $\mu=0$ [Fig. \ref{zes}(a)], while the number is $N_2$ for $\mu \ne 0$ [Fig. \ref{zes}(b)].
The system is gapless when $\Delta=0$ and $2 t |\eta| < \mu$:
See the horizontal black line in Fig. \ref{zes}(b).

\begin{figure}[t]
\centerline{\includegraphics[width=0.3\textwidth]{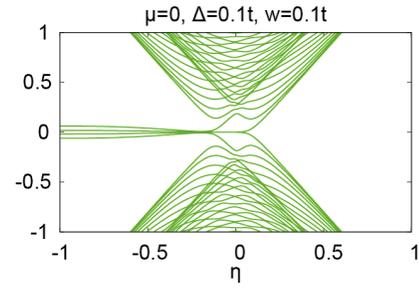}}
\caption{(Color online) Energy spectrum of the finite chain with disorder as a function of $\eta$, which corresponds to Fig. \ref{ene}(a). The distribution of the local disorder is fixed for all $\eta$. The zero-energy states in the SSH- (Kitaev-) like states are sensitive (robust) against the disorder. We have taken $L=64, \mu=0, \Delta=0.1t$, and $w=0.1t$.}
\label{disorder}
\end{figure}

We investigate the effect of the local disorder. We assume an onsite random potential, which will be relevant in experimetal realization.
We show the energy spectrum corresponding to Fig. \ref{ene}(a) by including the onsite random potential in Fig. \ref{disorder}, where we add the uniformly distributed random potential in $[-w,w]$.
The onsite random potential breaks the sublattice symmetry, while the particle-hole symmetry of the superconductivity is not broken.
Therefore, the SSH-like zero-energy states split, while the Kitaev-like zero energy is robust.
The effect of the disorder is common in all the parts of this paper, i.e., the SSH-like phase is sensitive and Kitaev-like phase is robust against the disorder.

\section{Differential Conductance} \label{sec:cnd}

We calculate the differential conductance of the normal metal/superconductor (NS) junction by means of the recursive Green's function method \cite{LeeFisher,FisherLee,He,Liu,Ii,Lewenkopf}.
We assume the normal lead has the same hopping $t$ and the chemical potential $\mu$ as the superconductor and there is no dimerization and superconducting order.
We define the hopping amplitude between the leads as $t_c$.
In order to obtain the differential conductance, we first calculate the surface Green's function of the semi-infinite superconductor numerically \cite{Sancho,Matsumoto,Umerski}.
In the Matsumoto--Shiba formalism \cite{Matsumoto}, the Nambu Green's function of the semi-infinite wire $\check{G}_{j,j'}$ is expressed by the Green's function of the bulk system $\check{G}^0_{j,j'}$:
\begin{equation}
\check{G}_{j,j'} = \check{G}^0_{j,j'} - \check{G}^0_{j,0} \left( \check{G}^0_{0,0} \right)^{-1} \check{G}^0_{0,j'}.
\end{equation}
We obtain $\check{G}^0_{j,j'}$ by performing Fourier transformation numerically for the $k$-space representation, which can be given analytically.
On the other hand, we give the analytic form of the surface Green's function for the semi-infinite normal lead \cite{Lewenkopf}.

Then, we construct the Green's function of the whole system by the following recursion relations.
Expressing the Green's function of the left (right) semi-infinite wire as $\check{G}_L (\check{G}_R)$ and the Green's function of the whole system as $\check{G}$,
\begin{align}
&\check{G}_{L,j,j}^{-1} = \check{g}_j^{-1} - \check{H}_{j,j-1} \check{G}_{L,j-1,j-1} \check{H}_{j-1,j}, \\
&\check{G}_{R,j,j}^{-1} = \check{g}_j^{-1} - \check{H}_{j,j+1} \check{G}_{R,j+1,j+1} \check{H}_{j+1,j}, \\
&\check{G}_{j,j}^{-1} = \check{g}_j^{-1} - \check{H}_{j,j-1} \check{G}_{L,j-1,j-1} \check{H}_{j-1,j} \notag \\
& \qquad \qquad \qquad \qquad - \check{H}_{j,j+1} \check{G}_{R,j+1,j+1} \check{H}_{j+1,j}, \\
&\check{G}_{j,j+1} = \check{G}_{j,j} \check{H}_{j,j+1} \check{G}_{R,j+1,j+1}, \\
&\check{G}_{j+1,j} = \check{G}_{j+1,j+1} \check{H}_{j+1,j} \check{G}_{L,j,j}, \\
&\check{g}_j^{-1} = E - \check{H}_j,
\end{align}
where $E$ is the energy, $\check{H}_j$ is the onsite Hamiltonian, $\check{H}_{j,j'}$ is the hopping between sites $j,j'$.
We obtain the retarded Green's function $\check{G}^\text{R}$ by replacing $E$ with $E + i \varepsilon$, where $\varepsilon$ is an infinitesimal factor.

After that, we calculate the differential conductance by the Lee-Fisher formula \cite{LeeFisher,Ii}
\begin{align}
	G=\frac{2e^2}{h} \text{Tr} \left[ \right. P_\text{e} \left( \right. &\check{G}''_{j,j+1} \check{G}''_{j,j+1} + \check{G}''_{j+1,j} \check{G}''_{j+1,j} \notag \\
& - \check{G}''_{j,j} \check{G}''_{j+1,j+1} - \check{G}''_{j+1,j+1} \check{G}''_{j,j} \left. \right) \left. \right],
\end{align}
where $\check{G}''_{j,j'} \equiv \text{Im} \check{G}^\text{R}_{j,j'}$, and $P_\text{e} \equiv \left( 1+\tau_z \right)/2$ is the projection operator onto the particle subspace.
We can choose an arbitrary $j$ in the normal region due to the current conservation.

\begin{figure}[t]
\centerline{\includegraphics[width=0.5\textwidth]{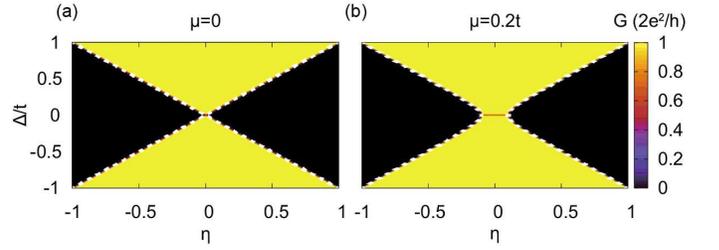}}
\caption{(Color online) Zero-bias differential conductance as a function of $\eta$ and $\Delta/t$ for (a) sublattice symmetric case ($\mu=0$) and (b) sublattice asymmetric case ($\mu=0.2t$). The white broken lines denote the phase boundaries. We have set $\varepsilon=0.001t$.}
\label{cnd}
\end{figure}

It is found that the zero-bias differential conductance corresponds not to $N_1$ but to the absolute value of $N_2$, i.e., the chiral index of Majorana fermions.
It is consistent with the fact that the zero-bias differential conductance in the NS junction takes nonzero values only when Majorana fermions exist.
Namely, the SSH-like zero-energy states do not contribute to the zero-bias differential conductance.
For nonzero $N_{2}$, the resulting conductance has a zero-bias peak. The magnitude of $G$ at zero-bias voltage is $2e^{2}/h$ reflecting on the perfect resonance via the zero-energy Andreev bound state as a Majorana fermion \cite{Tanaka1,KT,Bolech,Law,Wimmer,Fidkowski,Golub}.
We note that the quantized conductance does not depend on the coupling between the leads $t_c$.
When $\Delta=0$ and $2 t |\eta| < \mu$, a finite conductance whose magnitude is smaller than $e^2/h$ exists because the system is gapless, as is seen Fig. \ref{cnd}.
This conductance depends strongly on $t_c$, which is different from the quantized differential conductance in the topological region.

\section{Domain-Wall Soliton and Zero-Energy Modes} \label{sec:ldos}

We have so far analyzed fermion excitations around the ground-state configuration of the SSH model.
As is well known, a prominent feature of the SSH model is the existence of a soliton solution.
To discuss a soliton solution, it is necessary to include the kinetic and potential terms for the dimerization $\eta$ into the Hamiltonian (\ref{KSSHHamiltonian}). Such terms are summarized as
\begin{equation}
H_{\eta}=\sum_j\left\{\frac{(\hbar\dot{\eta_j})^2}{2M} + K(\eta_j-\eta_{j+1})^2+\lambda[(\eta_j)^2-(\eta)^2]^2\right\},
\end{equation}
where $M$, $K$, and $\lambda$ are constant parameters.
The ground-state solution is obviously given by $\eta_j=\eta$.
The soliton solution is given by
\begin{equation}
\eta_j = \eta \tanh \left[ \left( j - j_0 \right) a / \xi \right],
\end{equation}
where $\xi$ is the width of the soliton and $j_0$ is the center site index.

\begin{figure}[t]
\centerline{\includegraphics[width=0.5\textwidth]{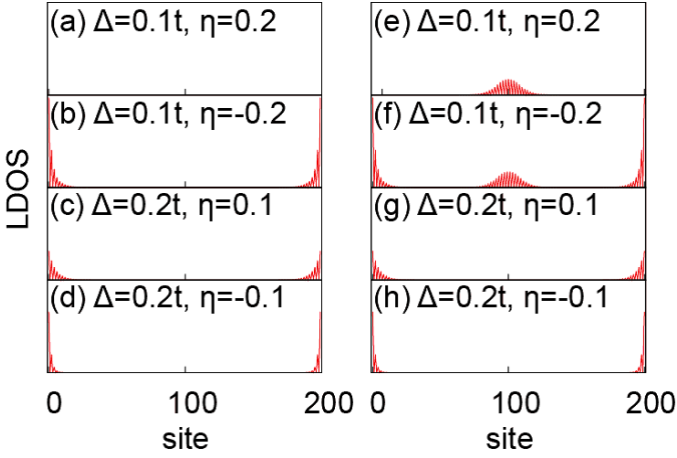}}
\caption{(Color online) LDOS at zero energy as a function of the position. In (a)--(d), the dimerization is constant and in (e)--(h), the soliton exists at the center. (a), (e) SSH-like trivial phase. (b), (f) SSH-like topological phase. (c), (d), (g), (h) Kitaev-like topological phase. In (e), (f), there are states around the soliton. The number of sites is $L=200$ in (a)--(d) and $L=201$ in (e)--(h). We have set $\mu=0, \xi=8a$, and $\varepsilon=0.001t$.}
\label{LDOS}
\end{figure}

We proceed to investigate fermion excitations in the presence of a soliton.
To demonstrate fermion excitations, we investigate the LDOS.
It is given by
\begin{equation}
\rho \left( E,j \right) = -\frac{1}{\pi} \text{Im} G^\text{R}(E,j,j)
\end{equation}
in terms of the retarded Green's function in the Nambu space,
\begin{equation}
\check{G}^\text{R}(E,j,j')=\left( \frac{1}{E-\check{H}+i \varepsilon} \right)_{j,j'} =
\begin{pmatrix}
G^\text{R} & F^\text{R} \\
\tilde{F}^\text{R} & \tilde{G}^\text{R}
\end{pmatrix}
.
\end{equation}
We show the LDOS at the zero energy without a soliton in Figs. \ref{LDOS}(a)--\ref{LDOS}(d).
In Fig. \ref{LDOS}(a), there is no state because the system is in the SSH-like trivial phase.
In Fig. \ref{LDOS}(b), there are edge states because the system is in the SSH-like topological phase.
In Figs. \ref{LDOS}(c) and \ref{LDOS}(d), there are Majorana zero-energy states because the system is in the Kitaev-like topological phase.

Now, we introduce a domain-wall (DW) soliton in dimerization.
We show the LDOS at the zero energy in Figs. \ref{LDOS}(e)--\ref{LDOS}(h).
In addition to the edge states, there are states localized around the soliton in the SSH-like phase [Figs. \ref{LDOS}(e) and \ref{LDOS}(f)].

\begin{figure}[t]
\centerline{\includegraphics[width=0.5\textwidth]{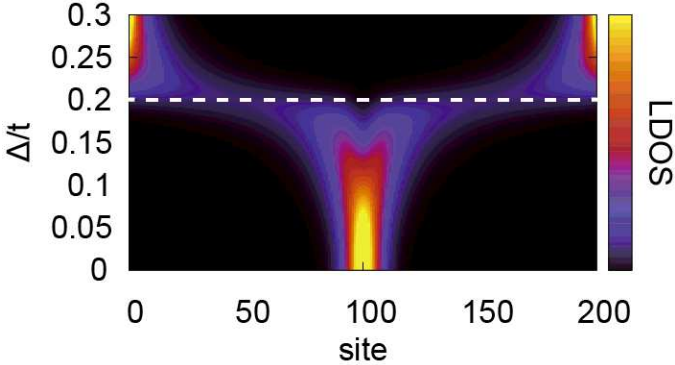}}
\caption{(Color online) Color plot of the LDOS as a function of the position and $\Delta/t$. For $\eta=0.2$, the system is in the SSH-like phase for $\Delta/t < 0.2$ and in the Kitaev-like phase for $\Delta/t > 0.2$. The broken line represents the transition point $\Delta/t = \eta$. We observe the precursor of the phase transition, i.e., splitting of the states at the soliton, in the SSH-like phase. We have set $L=201, \mu=0, \eta=0.2, \xi=8a$, and $\varepsilon=0.001t$.}
\label{LDOS_color}
\end{figure}

We also illustrate the LDOS as a function of the position and $\Delta/t$ in Fig. \ref{LDOS_color}, where we have set $\eta=0.2$.
When  $\Delta/t$ is smaller than $\eta$, there are states around the soliton. When $\Delta/t$ is equal to $\eta$, the gap closes and the states expand in the whole system. When $\Delta/t$ is larger than $\eta$, the soliton state disappears and a pair of Majorana fermions appears at the edges.
We find that the LDOS splits near the phase boundary in the SSH-like phase, which we will investigate later.

We make a further investigation of the zero-energy modes in the presence of a soliton in the continuum theory of our hybrid system.
The continuum limit of the SSH model is known as the Takayama-Lin-Liu-Maki model \cite{TLM}.
We take such a limit of our hybrid model, and derive an analytic expression of a soliton state appearing in the SSH-like phase.
We also derive the wave function at zero energy and the local density of states analytically, which are in accord with the numerical results.
In the following, we focus on the case $\mu=0$.

We introduce the right mover $R_j$ and the left mover $L_j$ by
\begin{align}
c_j &= e^{i k_F j a}R_j - i  e^{-i k_F j a}L_j, \\
c_j^\dagger &= e^{-i k_F j a}R_j^\dagger + i  e^{i k_F j a} L_j^\dagger,
\end{align}
where $k_F$ is the Fermi wave number.
We linearize the Hamiltonian by neglecting the high-frequency terms.
By introducing the spinor $\Psi^\dagger = \left( R^\dagger, L^\dagger, R, L \right)$, the result is written as
\begin{align}
& H = \frac{1}{2} \int dx \Psi \left( x \right)^\dagger \left( H_0 + H_\Delta \right) \Psi \left( x \right), \\
& H_0 = \hbar v_F \left[ - i \sigma_3 \partial_x + \frac{1}{a} \sigma_1 \tau_3 \eta \left( x \right) \right], \\
& H_\Delta = 2 \Delta \left[ - \sigma_2 \tau_2 + i a \tau_1 \eta \left( x \right) \partial_x \right], \label{HDelta}
\end{align}
where $v_F=2 t a/\hbar$ is the Fermi velocity, and
$\eta \left( x \right)$ is the space-depending dimerization.
We confirm the Hermiticity of the Hamiltonian, because the second term in Eq. (\ref{HDelta}) causes the terms such as $R^\dagger R^\dagger$ by the partial integration, and it vanishes due to the fermionic statistics.

We solve the eigenequation of the Hamiltonian.
The solution is given in the Appendix.
In particular, we take the soliton solution of the dimerization, $\eta \left( x \right) = \eta \tanh \frac{x}{\xi}$.
We set $\eta, \Delta > 0$ without loss of generality.
For the zero-energy solutions, the orthogonalized eigenfunctions are
\begin{equation}
\begin{pmatrix}
u_R \\ u_L \\ v_R \\ v_L
\end{pmatrix}
=
\begin{pmatrix}
h_+ \\ -i h_+ \\ h_+ \\ i h_+
\end{pmatrix},\quad
i
\begin{pmatrix}
h_- \\ -i h_- \\ -h_- \\ -i h_-
\end{pmatrix},
\end{equation}
where we have defined\begin{equation}
h_\pm \left( x \right) \equiv \text{e}^{\pm A x / a} \left( \cosh \frac{x}{\xi} \mp \frac{a \eta}{\xi_\Delta} \sinh \frac{x}{\xi} \right)^{-B \xi/a},
\label{EqAAA}
\end{equation}
with
\begin{equation}
A \equiv \frac{ \left( 1-\eta^2 \right) a \xi_\Delta}{\xi_\Delta^2 - a^2 \eta^2}, \quad
B \equiv \frac{\xi_\Delta^2 - a^2}{\xi_\Delta^2 - a^2 \eta^2}\eta, \quad
\xi_\Delta \equiv \frac{\hbar v_F}{2 \Delta}.
\end{equation}
The wave function is well defined only for $\xi_\Delta |\eta| > a$, that is, in the SSH-like phase.
We can check that the Majorana condition of the wave functions ($u_R^*=v_R, u_L^*=v_L$) is satisfied.
The peaks of the wave function amplitude locate at
\begin{equation}
x_\pm = \pm \xi \tanh^{-1} \frac{a}{\xi_\Delta \eta},
\end{equation}
whose amplitudes increase with $a/\xi_\Delta \eta$ and diverge at the phase boundary $\xi_\Delta \eta = a$.
$h_+$ and $h_-$ lean to the $x > 0$ region and the $x < 0$ region, respectively.
Namely, the two Majorana fermions split into right and left sides.

\begin{figure}[t]
\centerline{\includegraphics[width=0.5\textwidth]{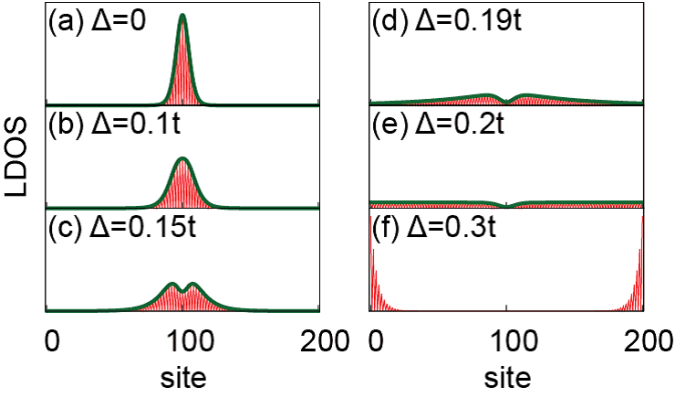}}
\caption{(Color online) LDOS obtained numerically (red line) and analytically (green line) for the various $\Delta$. The horizontal axis is the position and the vertical axis is the LDOS. The system is in the SSH-like phase for $\Delta < 0.2t$ and in the Kitaev-like phase for $\Delta > 0.2t$. We observe the precursor of the phase transition, i.e., splitting of the states at the soliton, in the SSH-like phase. We have set $L=201, \mu=0, \eta=0.2, \xi=8a$, and $\varepsilon=0.001t$.}
\label{LDOS_split}
\end{figure}

\begin{figure}[t]
\centerline{\includegraphics[width=0.5\textwidth]{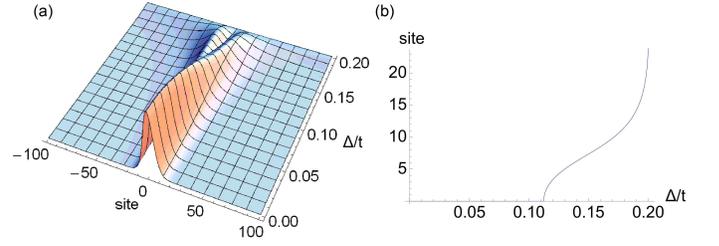}}
\caption{(Color online) (a) Analytic form of the LDOS as a function of the position and the $\Delta/t$. (b) The position of the LDOS peak as a function of $\Delta/t$. In both of the figures, we have set $\mu=0, \eta = 0.2$, and $\xi = 8a$.}
\label{LDOS_analytic}
\end{figure}

The LDOS at zero energy is given by
\begin{equation}
\rho \left( x , E=0 \right) \propto |h_+|^2 + |h_-|^2.
\end{equation}
In Fig. \ref{LDOS_split}, we show both the analytical result based on this formula and the numerical results based on the tight-binding model.
They fit very well, where we have shown the envelope function derived in analytic form.
We also show the analytic result as a function of the position and $\Delta/t$ in Fig. \ref{LDOS_analytic}(a).
We make an interesting observation.
When $\eta$ and $\Delta/t$ are compatible, the LDOS around the soliton split, which never occurs without the superconducting pairing.
It is regarded as a precursor of the topological phase transition, where the fermion at the soliton splits into two Majorana fermions.
We illustrate the position of the LDOS peak calculated numerically in Fig. \ref{LDOS_analytic}(b).
When $\Delta$ is sufficiently small, the LDOS peak locates at the center of the soliton.
However, at a certain critical point $\Delta_\text{c}$, the LDOS peak suddenly splits, and finally the position diverges at $\xi_\Delta \eta = a$.
In order to investigate the critical point, we expand the LDOS around $x=0$,
\begin{equation}
\rho \left( x, E=0 \right) \propto 2 + 2 \frac{ a^2 \eta + 2 a \xi - \xi_\Delta^2 \eta}{a \xi_\Delta^2 \xi} x^2 + O \left( x^4 \right).
\end{equation}
The critical value $\Delta_\text{c}$ is derived by the condition that the second term vanishes,
\begin{equation}
\Delta_\text{c} = \frac{\hbar v_F}{2 a} \sqrt{\frac{1}{1+2\xi/a \eta}}.
\end{equation}
It yields $\Delta_\text{c}=0.111\dots$ for $\eta=0.2$ and $\xi=8a$ as in Fig. \ref{LDOS_analytic}, which agrees well with the numerical result.

\section{Odd-frequency pairing} \label{sec:odd}
In order to understand the spatial dependence of the 
LDOS, it is useful to look at the symmetry of the Cooper pair. 
For this purpose, we calculate the Matsubara Green's function:
\begin{equation}
\check{G} \left( \omega_n, j, j' \right) = \left( \frac{1}{i \omega_n - \check{H}} \right)_{j,j'} =
\begin{pmatrix}
G & F \\
\tilde{F} & \tilde{G}
\end{pmatrix}
,
\end{equation}
in the Nambu space.
Owing to the Fermi--Dirac statistics, 
$F(\omega_n,j,j')=-F(-\omega_n,j',j)$ and $\tilde{F}(\omega_n,j,j')=-\tilde{F}(-\omega_n,j',j)$ are satisfied.
As regards the symmetry of the frequency, there are two possibilities: 
(i) $F(\omega_n,j,j')=F(-\omega_n,j,j')$ [$\tilde{F}(\omega_n,j,j')=\tilde{F}(-\omega_n,j,j')$] 
and 
(ii) $F(\omega_n,j,j')=-F(-\omega_n,j,j')$ [$\tilde{F}(\omega_n,j,j')=-\tilde{F}(-\omega_n,j,j')$]. 
The former and the latter cases correspond to even- and odd-frequency pairing amplitudes, respectively.
As for the exchange of $j$ and $j'$, the former one is odd parity and the latter the even parity.
In the inhomogeneous superconducting systems, 
like junctions or near the surface, translational
symmetry is broken. Then, the parity of the
Cooper pair is no more a good quantum number
and the mixed parity state can be realized.
If the symmetry of the bulk superconductor is  even (odd) parity, 
odd-frequency pairing with odd (even) parity is 
induced near the interface or the surface \cite{Tanaka2,Eschrig}. 
In the presence of the zero-energy surface Andreev bound state, it is known that the magnitude of the induced odd-frequency pairing amplitude is hugely enhanced \cite{Tanaka3,TSN} near the surface and is proportional to the inverse of $\omega_n$. Recent studies in the one-dimensional topological 
superconducting state in nanowire shows that the odd-frequency pairing is 
always generated where Majorana fermion exists since Majorana fermion is a special type of zero-energy Andreev bound state \cite{Asano,Stanev}.
In this work, since we are considering the spinless model and the symmetry of the bulk pair potential is $p$-wave (odd-parity) even frequency, we can naturally expect the generation of the $s$-wave (even-parity) odd frequency pairing near the edges or the kink.
We can show that 
$s$-wave odd-frequency is purely imaginary 
and $p$-wave even-frequency pairing amplitudes is 
purely real \cite{Tanaka3}. 
Thus, we plot the 
imaginary part of $s$-wave odd-frequency and real part of $p$-wave even-frequency pairing amplitude around site $j$ given by  $f_{\text{odd}}=- \text{Im} F \left( \omega_n, j, j \right)$ and $f_{\text{even}}=- \text{Re} F \left( \omega_n, j, j+1 \right)$.

\begin{figure}[t]
\centerline{\includegraphics[width=0.5\textwidth]{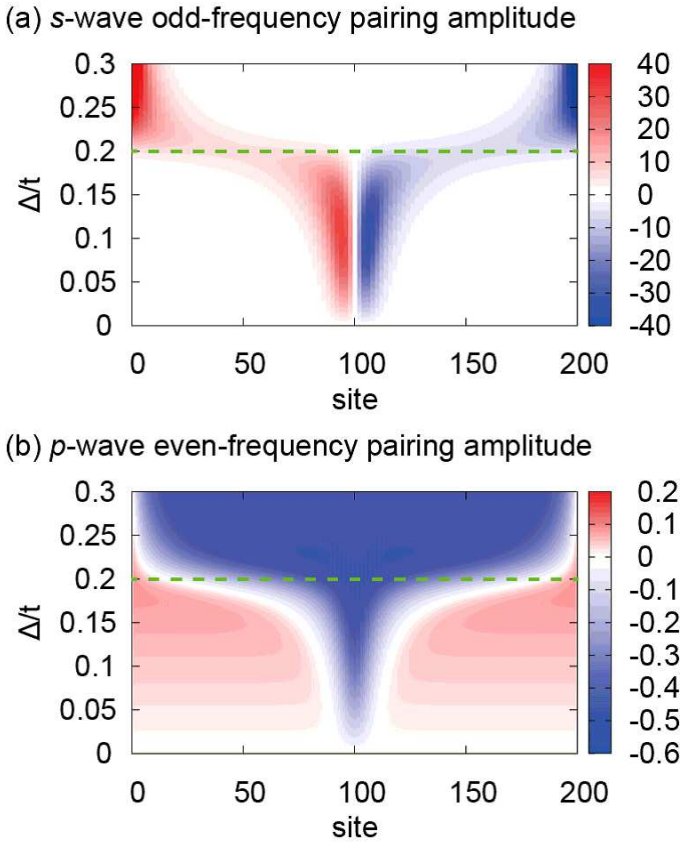}}
\caption{(Color online) Color plot of (a) $s$-wave odd-frequency $f_{\text{odd}}$ and (b) $p$-wave even-frequency pairing amplitude $f_{\text{even}}$ as a function of the position and $\Delta/t$. The broken lines represent the transition point $\Delta/t=\eta$. We have set $L=201, \mu=0, \eta = 0.2, \xi = 8a$, and $\omega_n=0.001t$.}
\label{pairing}
\end{figure}

In Fig. \ref{pairing}, $f_{\text{odd}}$ and $f_{\text{even}}$ 
are plotted for various sites and the magnitude of $\Delta/t$.
First, we focus on the $f_{\text{odd}}$.  
$f_{\text{odd}}$ is proportional to inverse of $\omega_{n}$, 
since it accompanies the zero-energy localized state. 
The qualitative feature of 
the absolute value of $f_{\text{odd}}$ 
is similar to that of LDOS as shown in Fig. \ref{LDOS_color}.
For small $\Delta/t$, 
the magnitude of $f_{\text{odd}}$ 
is enhanced  and has a sign change near the kink. 
On the other hand, $f_{\text{odd}}$ is enhanced at the edges
in the Kitaev-like phase with large magnitude of $\Delta/t$.
It is noted that $f_{\text{odd}}$ both at the left and right edges have opposite sign each other. 
This sign difference can induce anomalous Josephson coupling via proximity effect \cite{Tanaka4}. 
On the other hand, $p$-wave even-frequency pairing amplitude, $f_{\text{even}}$, 
stemming from the bulk state is enhanced 
where $s$-wave odd-frequency pairing is absent. 

\begin{figure}[t]
\centerline{\includegraphics[width=0.5\textwidth]{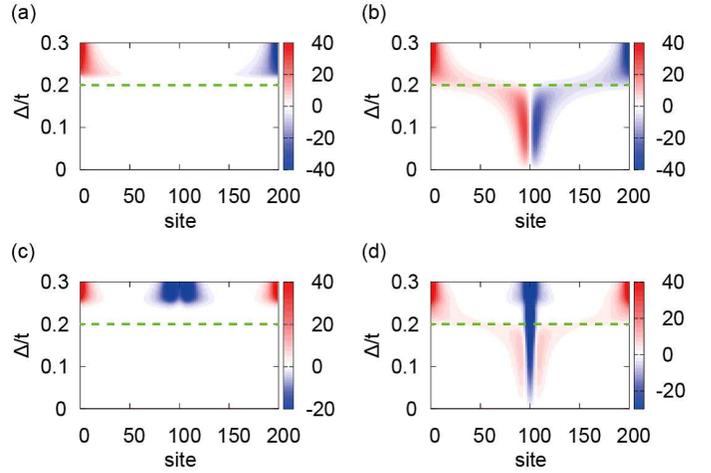}}
\caption{(Color online) Color plot of $s$-wave odd-frequency pairing amplitude $f_{\text{odd}}$. (a) No kink, (b) dimerization kink, (c) $p$-wave kink, and (d) dimerization kink and $p$-wave kink. The broken lines represent the transition point $\Delta/t=\eta$. We have set $L=200$ for (a) and (c), $L=201$ for (b) and (d), $\mu=0, \eta = 0.2, \xi = 8a$, and $\omega_n=0.001t$.}
\label{pairing_pattern}
\end{figure}

In order to understand the spatial dependence of 
$f_{\text{odd}}$ with dimerization kink in detail, 
we plot different configurations of dimerization and $p$-wave 
pair potential. 
As seen from the case with no dimerization kink 
[Fig. \ref{pairing_pattern}(a)], 
$f_{\text{odd}}$ only appears near the edge for $\Delta/t > 0.2$, i.e., Kitaev-like phase with 
Majorana fermion \cite{Asano}. 
$f_{\text{odd}}$ is generated in Kitaev-like phase as a Majorana fermion is localized at the edge state. 
The spatial dependence near the edges is similar to that in Figs. \ref{pairing}(a) and \ref{pairing_pattern}(b) (these two are identical). 
In the presence of the kink in the $p$-wave pair potential, i.e., the $p$-wave pairing changes the sign at the center of the chain [Fig. \ref{pairing_pattern}(c)], 
$f_{\text{odd}}$ appears for large magnitude of $\Delta$ with $\Delta/t > 0.25$, because of the finite-size effect. 
In this case, $f_{\text{odd}}$ is localized both near the kink and edges. 
$f_{\text{odd}}$ changes sign two times as a function of the site 
index $j$ and has the same sign  at left and right edges.  
Next, we consider the case 
with both dimerization and $p$-wave kinks, 
where the positions of kink are just the center of the chain. 
As seen from [Fig. \ref{pairing_pattern}(d)], 
$f_{\text{odd}}$ also changes sign two times as a function of $j$.  
For small magnitude of $\Delta/t$ with SSH-like phase, 
$f_{\text{odd}}$ is localized near the kink. 
By contrast to the case with dimerization kink,  
$f_{\text{odd}}$ is symmetric around the 
kink and  changes sign twice. 

\begin{figure}[t]
\centerline{\includegraphics[width=0.5\textwidth]{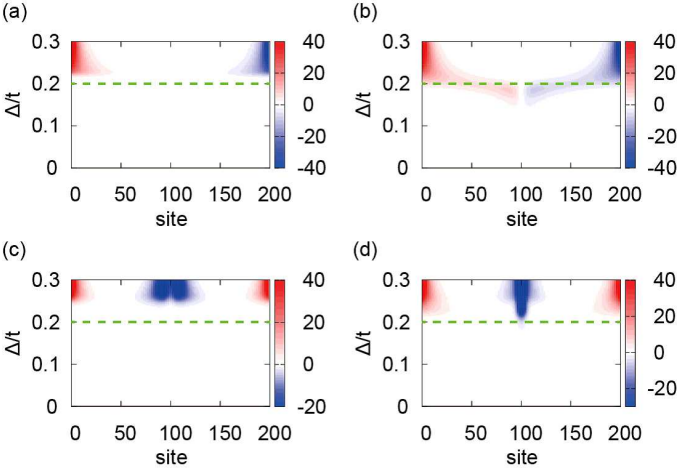}}
\caption{(Color online) Color plot of $s$-wave odd-frequency pairing amplitude $f_{\text{odd}}$ with the disorder. (a) No kink, (b) dimerization kink, (c) $p$-wave kink, and (d) dimerization kink and $p$-wave kink. The broken lines represent the transition point $\Delta/t=\eta$. We have set $L=200$ for (a) and (c), $L=201$ for (b) and (d), $\mu=0, \eta = 0.2, \xi = 8a$, and $\omega_n=0.001t$.}
\label{pairing_pattern_disorder}
\end{figure}

Finally, we consider the impurity effect on the $s$-wave odd-frequency pairing.
In Figs. \ref{pairing_pattern_disorder}, we plot the corresponding plot of the spatial dependence of $f_{\text{odd}}$ in the presence of the disorder.
(There is a one-to-one correspondence between Figs. \ref{pairing_pattern} and \ref{pairing_pattern_disorder}.)
In Fig. \ref{pairing_pattern_disorder}(b) for the case of dimerization kink, $f_{\text{odd}}$ near the kink existing in the SSH-like phase with $\Delta/t<0.2$ [see Fig. \ref{pairing_pattern}(b)] almost disappears. 
Since $f_{\text{odd}}$ has a sign change around the kink, it is expected that $f_{\text{odd}}$ is fragile against the ``pair annihilation'' of the positive and negative odd-frequency pairings due to the mixing caused by the disorder. 
On the other hand, $f_{\text{odd}}$ localized at the edges in the Kitaev-like phase is robust against the disorder.
This feature means that the zero-energy states in the SSH- (Kitaev-) like phase are sensitive (robust) against the disorder. 
We have also calculated spatial dependence of $f_{\text{odd}}$ for other three cases with including the disorder: (1) no kink [Fig. \ref{pairing_pattern_disorder}(a)], (2) $p$-wave kink [Fig. \ref{pairing_pattern_disorder}(c)], and (3) dimerization kink and $p$-wave kink [Fig. \ref{pairing_pattern_disorder}(d)].
In all of these cases, localized $f_{\text{odd}}$ near the edges in the Kitaev-like phase is robust against the disorder. 
On the other hand, localized $f_{\text{odd}}$ near the kink in the SSH-like phase is fragile against the disorder. 
The spatial dependence and sign of the $f_{\text{odd}}$ is important to understand the impurity effect on the Majorana fermion. 
As seen from these features, focusing on $s$-wave odd-frequency pairing is useful to understand the background of the physics of the Majorana fermion, especially the disorder effect.

\section{Conclusion and Discussion} \label{sec:conclusion}
In this paper, we have investigated the hybrid model comprised of the SSH model and the Kitaev model,
keeping a physical picture of polyacetylene with $p$-wave superconducting pairing in mind.
We have found that the system belongs to either the SSH-like or Kitaev-like phase depending on the relative strength between the dimerization and the superconducting pairing.
We have found there are two types of particle-hole symmetries due to the sublattice symmetry (SSH like) or the superconductivity (Kitaev like).
We can define the $\mathbb{Z}$ index for each symmetry:
$N_1$ corresponds to the number of zero-energy states at $\mu=0$, while
$N_2$ corresponds to the zero-bias differential conductance for arbitrary values of $\mu$.
We have found the splitting of the states around a soliton when the superconducting pairing is comparable to the dimerization strength.
It is regarded as the splitting of the fermion into the two Majorana fermions, which is a precursor of the topological phase transition.
We have found $s$-wave odd-frequency pairing amplitude is strongly enhanced around the splitted states.
The model may be realized in the organic superconductor or by putting a polyacetylene on an intrinsic $p$-wave superconductor such as Sr$_2$RuO$_4$.
There are also possibilities to realize the model by using $s$-wave superconductor and engineering the Rashba spin-orbit interaction by placing micro-magnets \cite{Braunecker,Choy,Kjaergaard,Klinovaja1,Vazifeh,Klinovaja2,Braunecker2}, or quantum-dot array \cite{Fulga}.

{\it Note added in proof.} Recently, we became aware of papers on the similar topic \cite{Sticlet,Klinovaja3,Rainis}.

\section*{Acknowledgements}

This work was supported by a Grantin-Aid for Scientific Research (S) 
(Grant No. 24224009 and 25400317);
the Funding Program for World-Leading Innovative RD on
Science and Technology (FIRST Program); 
the Strategic International Cooperative Program (Joint Research Type) from the
Japan Science and Technology Agency; 
and Innovative Areas ``Topological Quantum Phenomena'' (Grant No. 22103005)
from the Ministry of Education, Culture, Sports, Science, and
Technology of Japan.

\appendix

\section{Continuum model}
In this appendix, we derive the zero-energy solution (\ref{EqAAA}) based on the continuum model,
\begin{align}
& H = \frac{1}{2} \int dx \Psi \left( x \right)^\dagger \left( H_0 + H_\Delta \right) \Psi \left( x \right), \\
& H_0 = \hbar v_F \left[ - i \sigma_3 \partial_x + \frac{1}{a} \sigma_1 \tau_3 \eta \left( x \right) \right], \\
& H_\Delta = 2 \Delta \left[ - \sigma_2 \tau_2 + i a \tau_1 \eta \left( x \right) \partial_x \right],
\end{align}
where $\Psi^\dagger = \left( R^\dagger, L^\dagger, R, L \right)$ and $v_F=2ta/\hbar$ is the Fermi velocity.
We define $\xi_\Delta \equiv \hbar v_F / 2 \Delta$, which has the same order as the superconducting coherence length.
The eigenequations for zero energy are
\begin{align}
&\xi_\Delta \left[ - i a \partial_x u_R + \eta \left( x \right) u_L \right] + a \left[ i a \eta \left( x \right) \partial_x v_R + v_L \right] = 0, \\
&\xi_\Delta \left[ i a \partial_x u_L + \eta \left( x \right) u_R \right] + a \left[ i a \eta \left( x \right) \partial_x v_L - v_R \right] = 0, \\
&\xi_\Delta \left[ - i a \partial_x v_R - \eta \left( x \right) v_L \right] + a \left[ i a \eta \left( x \right) \partial_x u_R - u_L \right] = 0, \\
&\xi_\Delta \left[ i a \partial_x v_L - \eta \left( x \right) v_R \right] + a \left[ i a \eta \left( x \right) \partial_x u_L + u_R \right] = 0
.
\end{align}
We define $f_\pm \equiv u_R \pm i u_L, g_\pm \equiv v_R \pm i v_L$.
Then, these equations are grouped into two set of equations,
\begin{align}
\begin{cases}
\xi_\Delta \left[ a \partial_x - \eta \left( x \right) \right] f_- - a \left[ a \eta \left( x \right) \partial_x - 1 \right] g_+ &= 0 \\
\xi_\Delta \left[ a \partial_x - \eta \left( x \right) \right] g_+ - a \left[ a \eta \left( x \right) \partial_x - 1 \right] f_- &= 0 
\end{cases}
\end{align}
and
\begin{align}
\begin{cases}
\xi_\Delta \left[ a \partial_x + \eta \left( x \right) \right] f_+ - a \left[ a \eta \left( x \right) \partial_x + 1 \right] g_- &= 0 \\
\xi_\Delta \left[ a \partial_x + \eta \left( x \right) \right] g_- - a \left[ a \eta\left( x \right)  \partial_x + 1 \right] f_+ &= 0
\end{cases}
.
\end{align}
Furthermore, we decouple them into independent equations,
\begin{align}
&\left\{ a \left[ \xi_\Delta \mp a \eta \left( x \right) \right] \partial_x - \left[ \xi_\Delta \eta \left( x \right) \mp a \right] \right\} \left( f_- \pm g_+ \right) = 0, \\
&\left\{ a \left[ \xi_\Delta \mp a \eta \left( x \right) \right] \partial_x + \left[ \xi_\Delta \eta \left( x \right) \mp a \right] \right\} \left( f_+ \pm g_- \right) = 0	
.
\end{align}
This can be solved for general $\eta \left( x \right)$ as
\begin{align}
f_- \pm g_+ &= \exp \left[ + \frac{1}{a} \int^x \frac{\xi_\Delta \eta \left( x' \right) \mp a}{\xi_\Delta  \mp a \eta \left( x' \right)}dx' \right] \\[1mm]
f_+ \pm g_- &= \exp \left[ - \frac{1}{a} \int^x \frac{\xi_\Delta \eta \left( x' \right) \mp a}{\xi_\Delta  \mp a \eta \left( x' \right)}dx' \right]
.
\end{align}
Here we substitute the soliton of the dimerization $\eta \left( x \right) = \eta \tanh \frac{x}{\xi}$ to find
\begin{align}
f_- \pm g_+ &= \text{e}^{\mp A x / a} \left( \cosh \frac{x}{\xi} \mp \frac{a \eta}{\xi_\Delta} \sinh \frac{x}{\xi} \right)^{+ B \xi / a}, \label{EqA1}\\
f_+ \pm g_- &= \text{e}^{\pm A x / a} \left( \cosh \frac{x}{\xi} \mp \frac{a \eta}{\xi_\Delta} \sinh \frac{x}{\xi} \right)^{- B \xi / a}\label{EqA2}, 
\end{align}
with
\begin{equation}
A \equiv \frac{ \left( 1-\eta^2 \right) a \xi_\Delta}{\xi_\Delta^2 - a^2 \eta^2}, \quad
B \equiv \frac{\xi_\Delta^2 - a^2}{\xi_\Delta^2 - a^2 \eta^2}\eta.
\end{equation}
We can set $\Delta, \eta > 0$ without loss of generality.
Then, because Eq. (\ref{EqA1}) diverges at $x \rightarrow \pm \infty$, these coefficients have to be zero.
Regarding Eq. (\ref{EqA2}), the dominant factor at $x \rightarrow \pm \infty$ is
\begin{equation}
\exp \left( A - B \right) = \exp \left[ \frac{a - \xi_\Delta \eta}{\xi_\Delta - a \eta} |x| / a \right].
\end{equation}
Therefore, the condition $\xi_\Delta \eta> a$ has to be satisfied, which accords with the SSH-like phase of the tight-binding model.
Under this condition,  the orthogonalized eigenfunctions are
\begin{equation}
\begin{pmatrix}
u_R \\ u_L \\ v_R \\ v_L
\end{pmatrix}
=
\begin{pmatrix}
h_+ \\ -i h_+ \\ h_+ \\ i h_+
\end{pmatrix},\quad
i
\begin{pmatrix}
h_- \\ -i h_- \\ -h_- \\ -i h_-
\end{pmatrix},
\end{equation}
where we have defined
\begin{equation}
h_\pm \left( x \right) \equiv \text{e}^{\pm A x / a} \left( \cosh \frac{x}{\xi} \mp \frac{a \eta}{\xi_\Delta} \sinh \frac{x}{\xi} \right)^{-B \xi / a}.
\end{equation}
Hence we have derived Eq. (\ref{EqAAA}).
Without the superconducting pairing, we obtain
\begin{equation}
h_\pm \left( x \right) = \left( \cosh \frac{x}{\xi} \right)^{- \eta \xi / a}.
\end{equation}
which is well known in literature \cite{Niemi,JR}.

\end{document}